\documentclass[sigconf,authorversion,nonacm]{acmart} 
\usepackage{amsmath,amsfonts,mathtools}
\usepackage{subcaption}
\usepackage{comment}
\usepackage{array}
\usepackage{xcolor}
\usepackage{tikz}
\usepackage{enumitem}
\usepackage[font=small,skip=0pt]{caption}
\usepackage[normalem]{ulem}

\newcommand{\edll}[1]{{\color{green} Ed's Lesson Learned: #1}}

\newcommand{\edc}[1]{{\color{orange}}}
\newcommand{\pp}[1]{{\color{red} Purpose of paragraph: #1}}
\newcommand{\qc}[1]{{\color{orange} Questions/Concerns: #1}}

\begin{document}

\title{Application Performance Modeling via Tensor Completion}

\author{Edward Hutter}
\email{hutter2@illinois.edu}
\affiliation{%
  \department{Department of Computer Science}
  \institution{University of Illinois at Urbana-Champaign}
  \city{Champaign}
  \state{IL}
  \country{USA}
}
\author{Edgar Solomonik}
\email{solomon2@illinois.edu}
\affiliation{%
  \department{Department of Computer Science}
  \institution{University of Illinois at Urbana-Champaign}
  \city{Champaign}
  \state{IL}
  \country{USA}
}



\setlength{\abovedisplayskip}{4pt}
\setlength{\belowdisplayskip}{4pt}
\setlength{\abovedisplayshortskip}{4pt}
\setlength{\belowdisplayshortskip}{4pt}

\begin{abstract}
  Performance tuning, software/hardware co-design, and job scheduling are among the many tasks that rely on models to predict application performance.
  We propose and evaluate low-rank tensor decomposition for modeling application performance.
  We discretize the input and configuration domains of an application using regular grids.
  Application execution times mapped within grid-cells are averaged and represented by tensor elements.
  We show that low-rank canonical-polyadic (CP) tensor decomposition is effective in approximating these tensors.
  We further show that this decomposition enables accurate extrapolation of unobserved regions of an application's parameter space.
  We then employ tensor completion to optimize a CP decomposition given a sparse set of observed execution times.
  We consider alternative piecewise/grid-based models and supervised learning models for six applications and demonstrate that CP decomposition optimized using tensor completion offers higher prediction accuracy and memory-efficiency for high-dimensional performance modeling.
\end{abstract}

\maketitle


\section{Introduction}
    Application performance depends on 
    both configuration parameters (e.g., block sizes, processor grid topology) and architectural parameters (e.g., processes-per-node, hyper-thread count) for a given set of inputs.
    Complex interactions among these parameters~\cite{vuduc2004statistical,roy2021bliss} motivate observation of an application's full parameter space, the size of which has increased to account for growth in algorithmic complexity and architectural diversity.
    Tasks such as program optimization~\cite{petrini2003case,hoefler2011performance,calotoiu2013using}, optimal tuning parameter selection~\cite{liu2021gptune,roy2021bliss}, architectural design~\cite{lee2006accurate,lee2007methods}, dynamic load balancing, and machine allocation estimation
    (e.g., for weather forecasting) increasingly rely upon predictive models
    for accurate and efficient performance prediction.

    The demand for increasingly accurate performance prediction is reflected in the abundance of available performance modeling frameworks (e.g., ~\cite{calotoiu2016fast,tallent2014palm,copik2021extracting,choi2020end}). 
    Application performance models, which we survey in Section~\ref{sec:review}, are typically either (1) global (non-piecewise) models configured semi-analytically or automatically (such as via \textit{performance model normal form} ~\cite{calotoiu2013using,calotoiu2016fast}),
    (2) piecewise/ grid-based models, which discretize the modeling domain (such as \textit{sparse grid regression} (SGR)~\cite{pflueger10spatially}), or (3) alternative models configured by supervised learning methods (such as neural networks).
    We instead consider tensor decompositions, which generalize low-rank matrix factorizations to higher dimensions~\cite{kolda2009tensor} and are effective models for tensor completion~\cite{tomasi2005parafac,acar2011scalable}, the problem of approximating a tensor given a subset of observed entries.
    Specifically, we employ tensor completion based on low-rank CP decomposition~\cite{hitchcock1927expression} for application performance modeling, and compare this model, which belongs to type (2), to representative models of each type.

    A CP decomposition models an application configuration $\mathbf{x}\equiv (x_{1},\ldots,x_{d})$'s execution time $f(\mathbf{x})$ as a linear combination of separable piecewise functions $m(\mathbf{x})=\sum_{r=1}^{R}\prod_{j=1}^{d}g_{r,j}(x_{j})$.
    A CP decomposition thus generalizes simple multilinear execution cost models.
    These models also provide systematic improvability (via increased rank) and achieve linear model size with tensor order (number of application parameters) for a fixed rank.
    We provide specific details on how we discretize an application's parameter space to construct a low-rank CP decomposition model in Section \ref{sec:low-rank-perf-model}.

    We evaluate this model in two settings: interpolation (training set sampled from same input and configuration domain as test set) and extrapolation (test set contains larger scale problems than training set).
    We consider an error metric $|\log(m(\mathbf{x})/f(\mathbf{x}))|$ that is independent of the scale of the execution times.
    To minimize this error, we consider two loss functions.
    First, we simply transform the input data and minimize $(\hat{m}(\mathbf{x})-\log(f(\mathbf{x})))^{2}$, resulting in model $m(\mathbf{x})=e^{\hat{m}(\mathbf{x})}$.
    Second, we consider loss function $\log(m(\mathbf{x})/f(\mathbf{x}))^2$, which requires enforcing positivity of an alternative model $m(\mathbf{x})$.
    The positivity of the latter model is leveraged by our proposed method for extrapolation, as it employs rank-1 factorizations that retain positivity.

    The most similar technique to CP decomposition studied in prior work for application performance modeling is \textit{sparse grid regression} (SGR)~\cite{pflueger10spatially,neumann2019sparse}.
    SGR uses a hierarchical reduced-size model of a regular grid containing performance measurements, whereas we first discretize an application's parameter space directly on a regular grid before formulating a CP decomposition of the corresponding tensor.
    Both SGR and CP decomposition can compress performance across high-dimensional modeling domains efficiently.
    Our experimental results in Section \ref{sec:eval} show that low-rank CP decomposition offers improved accuracy and scalability in high-dimensional domains, and further facilitates accurate performance extrapolation.

    Our experimental evaluation considers randomly sampled execution times of 6 benchmarks: QR factorization, matrix multiplication, MPI Broadcast, ExaFMM~\cite{yokota2013fmm}, AMG~\cite{richards2018quantitative}, and KRIPKE~\cite{kunen2015kripke} on the Stampede2 supercomputer~\cite{Stanzione:2017:SEX:3093338.3093385}.
    These applications have 2-12 parameters (including input parameters, configuration parameters, and architectural parameters), yielding tensors of corresponding order in our model.
    Our experiments consider 9 alternative methods: sparse grid regression, multi-layer perceptrons, random forests, gradient-boosting, extremely-randomized trees, Gaussian process regression, support vector machines, adaptive spline regression, and k-nearest neighbors.
    Among these methods, we find our tensor model to be the most effective in multiple metrics, including model accuracy and model size. 
    Our model is especially efficient in comparison to other models for applications with high-dimensional (many parameter) input and configuration domains.

    Our novel contributions towards application performance modeling are as follows:
    \begin{itemize}[noitemsep,topsep=0pt]
        \item a methodology for learning piecewise multilinear models via application of CP tensor decomposition (CPR),
        \item a technique for performance extrapolation using CP tensor decomposition,
        \item a software library for CPR that leverages open-source high-performance software for tensor completion\footnote{https://github.com/huttered40/cpr-perf-model},
        \item an evaluation of piecewise/grid-based models and alternative supervised learning methods using six applications,
        \item empirical evidence that CPR achieves highest prediction accuracy among existing models relative to model size.
    \end{itemize}

\section{Problem Statement}\label{sec:problem_statement}  
    Application performance models predict execution times given a configuration of application benchmark parameters.
    We formalize performance modeling and summarize quantitative techniques to assess performance models in this section.

    \subsection{Modeling Problem}\label{sec:problem_statement:overall}
    We consider the problem of predicting execution times of an application for any configuration (i.e., selection of parameter values for each of a pre-specified set of parameters), given execution times of an observed subset of an application's parameter space.
    Formally, the execution time of an application is represented by a latent multivariate function $f : \mathcal{X}\rightarrow \mathbb{R}^{+}$, which attributes positive real values to all configurations across its $d$-dimensional parameter space $X^{(1)}\times\ldots\times X^{(d)}\equiv\mathcal{X}$.
    Each configuration $(x_{1},\ldots,x_{d})\equiv\mathbf{x}\in\mathcal{X}$ is comprised of input parameters (e.g., matrix dimension), configuration parameters (e.g., block size), and/or architectural parameters (e.g., thread count).
    These parameters may be real, of integer type, or categorical (e.g., choice of solver).
    In the absence of a set of benchmark parameters, statistical techniques (e.g., clustering and correlation analysis~\cite{lee2006accurate,lee2007methods}), which we do not evaluate, can facilitate specification of configuration space $\mathcal{X}$.

    A performance model $m : \tilde{\mathcal{X}}\rightarrow \mathbb{R}$ maps a $d$-parameter configuration $\mathbf{x}\in\tilde{\mathcal{X}}$ to a real value $m(\mathbf{x})$, which represents a prediction of true execution time $f(\mathbf{x})$.
    The modeling domain $\tilde{\mathcal{X}}\subseteq X^{(1)}\times\ldots\times X^{(d)}$ is a subset of an application's configuration space $\mathcal{X}$, and is ascertained either from a set of previously executed configurations (i.e., training set) or by intuition.
    Configurations outside of the modeling domain $\mathbf{x}\notin\tilde{\mathcal{X}}$ induce extrapolation, while configurations within the modeling domain $\mathbf{x}\in\tilde{\mathcal{X}}$ induce interpolation.
    The prediction accuracy of a performance model should be robust to extrapolation and generalize effectively for interpolation.
    We evaluate both settings for model inference in Section \ref{sec:eval}.

    Predictive models for estimating application performance can be classified as analytic, learned (i.e., data-driven), or some combination therein.
    Analytic models rely on program analysis and/or expert knowledge of both an application and architecture, and decompose an application's execution time into execution times contributed by computational and communication kernels executed along its execution call paths~\cite{rauber2000modelling,kerbyson2001predictive,goldsmith2007measuring,hoefler2010toward,hoefler2011performance,tallent2014palm,friese2017generating}.
    Domain expertise and static source code analysis alone have been shown to achieve sufficient accuracy in modeling performance to facilitate optimal configuration selection for both non-distributed kernels~\cite{10.1145/2925987,li2019analytical,10.1145/3445814.3446759} and distributed kernels~\cite{nuriyev2021efficient,georganas2012communication}.
    We instead consider and evaluate learned models for estimating application performance that are configured solely using execution times of observed configurations.

    \subsection{Model Assessment}\label{sec:problem_statement:assessment}

    \begin{table}[t]
        \centering
        \begin{tabular}{||m{3.75em}|m{11em}|m{8.4em}||} 
            \hline
            Metric & Mathematical Expression & Error Expression\\ 
            \hline\hline
            MAPE & $\sum_{k=1}^{M} \frac{|m_{k}-y_{k}|}{y_{k}}$ &$\sum_{k=1}^{M}|\epsilon_{k}|$\\
            \hline
            MAE & $\sum_{k=1}^{M} |m_{k}-y_{k}|$ &$\sum_{k=1}^{M}|y_{k}\epsilon_{k}|$\\
            \hline
            MSE & $\sum_{k=1}^{M} (m_{k}-y_{k})^{2}$ &$\sum_{k=1}^{M}\left(y_{k}\epsilon_{k}\right)^{2}$\\
            \hline
            SMAPE & $2\sum_{k=1}^{M} \frac{|m_{k}-y_{k}|}{y_{k}+m_{k}}$ &$2\sum_{k=1}^{M}|\frac{\epsilon_{k}}{2+\epsilon_{k}}|$\\
            \hline
            LGMAPE & $\sum_{k=1}^{M} \log(\frac{|m_{k}-y_{k}|}{y_{k}})$ &$\sum_{k=1}^{M}\log(|\epsilon_{k}|)$\\
            \hline
            MLogQ & $\sum_{k=1}^{M} |\log(\frac{m_{k}}{y_{k}})|$ &$\sum_{k=1}^{M}|\frac{\epsilon_{k}}{1+\epsilon_{k}}|+\mathcal{O}(\epsilon_{k}^{2})$\\
            \hline
            MLogQ2 & $\sum_{k=1}^{M} \log^{2}(\frac{m_{k}}{y_{k}})$ &$\sum_{k=1}^{M}(\frac{\epsilon_{k}}{1+\epsilon_{k}})^{2}+\mathcal{O}(\epsilon_{k}^{4})$\\
            \hline
        \end{tabular}
        \caption{Error metrics and corresponding expressions (scaled by $M$) given model predictions $m$, true positive outputs $y$, and errors $\epsilon=m/y-1$. Both expressions are equivalent across rows 1-5, and equivalent to low-order Taylor approximation in $\epsilon$ across rows 6-7.}
        \label{error_metric_table}
    \end{table}
    
    Performance models may be assessed in terms of model size, optimization complexity, inference time, and/or the accuracy with which they predict application performance.
    Given a particular configuration of application parameters $(x_{1},\ldots,x_{d})\equiv \mathbf{x}$, execution time $y\equiv f(\mathbf{x})$ acquired by averaging a sufficient number of samples, and a model output $m\equiv m(\mathbf{x})$, relative prediction error is given by $|m-y|/y$.
    While relative error is more appropriate than absolute error for assessing performance predictions spanning multiple orders of magnitude (e.g., application evaluation over a range of inputs), it does not assign similar errors to mispredictions of a similar scale (i.e., $|\log(m/y)|$).
    Model selection via minimization of relative error exhibits bias towards under-prediction~\cite{tofallis2015better},
    which can misinform tasks that leverage model output (e.g., optimal parameter selection).

        To assess performance prediction accuracy, we adopt error metrics that assign equal error to under/over-predictions of the same scale (i.e., scale independence).
        In particular, we penalize model outputs $m_{k}=ay_{k}$ and $\tilde{m}_{k}=y_{k}/a$ equally for positive factor $a$.
        Consider a collection of measured configurations $\{(\mathbf{x}_{k},y_{k})\}_{k=1}^{M}$ and model outputs $\{(\mathbf{x}_{k},m_{k}))\}_{k=1}^{M}$.
        Among the aggregate error metrics listed in Table \ref{error_metric_table}, only the arithmetic means of the absolute log and log-squared accuracy ratios (MLogQ and MLogQ2, respectively) achieve scale independence~\cite{tofallis2015better,morley2018measures}.
        This table also demonstrates the equivalence of the symmetric mean and log-transformed geometric-mean relative errors (SMAPE and LGMAPE, respectively), and MLogQ to first-order Taylor approximation for small relative errors $\epsilon_{k}$ defined as $m_{k}=y_{k}(1+\epsilon_{k})$.
        We configure models to minimize MLogQ in Section \ref{sec:exp_methodology}, and describe alternative loss functions that target this in Sections \ref{sec:cpd_model_formulation:interp} and \ref{sec:cpd_model_formulation:extrap}.

\section{Multi-Parameter Performance Modeling}\label{sec:review}
    Application performance modeling uses model class selection and optimization to minimize an application-dependent error metric.
    Selection of a class of models reflects application characteristics (e.g., scaling behavior, discontinuities) identified analytically and/or learned using measurements of executed configurations.
    Following a selection of class-specific hyper-parameters, a candidate model is configured by minimizing an aggregate error metric on execution times of observed configurations.
    We review methods to construct performance models that address the modeling problem described in Section \ref{sec:problem_statement:overall}.

    \subsection{Global} \label{sec:review:global}
    Relationships between application benchmark parameters and execution times are expressed as linear combinations of predictor variables (i.e., non-piecewise functions parameterized on a subset of benchmark parameters), each of which has global support across the parameter space.
    These variables model interaction effects and nonlinear scaling behavior among subsets of parameters.
    Ordinary least-squares regression (OLS) is a quadratic optimization problem that estimates unknown parameters expressed linearly with respect to predictor variables by minimizing mean squared error (MSE).
    Use of OLS or ridge regression to optimize global models were among the first class of methods proposed for modeling application performance empirically~\cite{lee2006accurate,lee2007methods,goldsmith2007measuring,barnes2008regression} and remain widely used, especially in settings in which a limited number of configurations can be executed.

    Expert insight, statistical analysis, and automated search have each been proposed to find suitable predictor variables to instantiate interpretable global models with varying complexity.
    Adaptations to Lasso regression have been shown to effectively prune a large space of predictors~\cite{huang2010predicting}.
    Alternative methods to circumvent statistical analyses for predictor selection include symbolic regression~\cite{chenna2019multi} and search, which performs OLS repeatedly on candidates from canonical classes of small global models.
    In particular, the performance model normal form ~\cite{calotoiu2013using,calotoiu2016fast} expresses execution time as
    \begin{align}\label{eq:pmnf}
        m(\mathbf{x})=\sum_{r=1}^{R}\alpha_{r}\prod_{j=1}^{d}x_{j}^{v_{r,j}}\log^{w_{r,j}}(x_{j}),
    \end{align}
    which defines a search space, the size of which is predicated on user-specified sets of rational numbers $v,w$.
    Global models that utilize log-transformed benchmark parameters as predictors significantly reduce this predictor search space while retaining tolerable prediction accuracy~\cite{barnes2008regression}.
    
    \subsection{Piecewise/Grid}\label{sec:review:grid}
    Non-piecewise predictor variables can be insufficient for modeling complex performance behavior with high accuracy.
    Complex dependence of performance on application parameters arises even in simple programs due to e.g., memory misalignment, register spilling, and load imbalance.
    This behavior can be more accurately modeled as a linear combination of spline functions, i.e., tensor products of univariate functions defined piecewise by polynomials.
    These models introduce additional hyper-parameters, notably the selection of grid-points that discretize the modeling domain.
    Therefore, it is common to formulate global models, and only segment those parameters whose values correlate strongest with performance~\cite{lee2006accurate,lee2007methods}.

    Multilinear interpolation methods evaluate an application's execution time $f(\mathbf{x})$ on $d$-dimensional regular grids and model $f(\mathbf{x})$ as a weighted average of $2^{d}$ neighboring grid-points, each of which stores the measured execution time of a distinct configuration.
    For example, bilinear interpolation predicts a configuration $(x_{1},x_{2})$'s execution time as
    \begin{align*}
        m(\mathbf{x})=\sum_{a=i_{1}}^{i_{1}+1}\sum_{b=i_{2}}^{i_{2}+1} t_{a,b} \bigg(1-\frac{|x_{1}-x_{a}|}{x_{i_{1}+1}-x_{i_{1}}}\bigg)\bigg(1-\frac{|x_{2}-x_{b}|}{x_{i_{2}+1}-x_{i_{2}}}\bigg),
    \end{align*}
    where $x_{i_{j}}\le x_{j}< x_{i_{j}+1},\forall j\in\{1,2\}$ and $t_{a,b}$ denotes the execution time of grid-point $(a,b)$ on a two-dimensional regular grid.
    Multilinear interpolation is suitable if performance behavior within grid-cells is sufficiently smooth.
    However, this is rarely achieved without a significantly large number of observations.
    
    Multiple frameworks, including \textit{multivariate adaptive regression splines} (MARS)\cite{friedman1991multivariate} and \textit{sparse grid regression}~\cite{pflueger12spatially,pflueger10spatially}, instead compress regular grids by selecting a subset of grid-points and corresponding spline functions automatically.
    MARS recursively constructs products of univariate hinge functions $h(x_{i})=\{\text{c},\text{max}(0,x_{i}-c)\}$, each characterized by a parameter $x_{i}$ and a scalar position $c$ within its range.
    The selection of these parameters involves repeatedly searching across the parameter space and invoking OLS as candidate models are constructed and pruned.
    We observe in Section \ref{sec:eval} that MARS performance models~\cite{courtois2000using} achieve only a coarse discretization of high-dimensional modeling domains and often produce global models.
    
    Sparse grid regression (SGR) discretizes high-dimensional functions on anisotropic sparse grids.
    Sparse grids~\cite{yserentant1992hierarchical,smolyak1963quadrature,garcke2012sparse} are characterized by a hierarchical representation of piecewise-linear functions, the tensor-product of which yields piecewise $d$-linear basis functions.
    Only functions with sufficiently-large support across the modeling domain are retained.
    Sparse grids initially utilize $\mathcal{O}(2^{n}n^{d-1})$ grid-points, where $n$ is a user-specified discretization level.
    SGR models an application's execution time as a linear combination of $\mathcal{O}(2^{n}n^{d-1})$ multi-scale basis functions~\cite{neumann2019sparse} and offers automated spatially-adaptive grid refinement within the support of a specified number of basis functions~\cite{neumann2019sparse}.
    Our methodology circumvents grid refinement and compresses regular grids, which feature $\mathcal{O}(2^{nd})$ total grid-points, using rank-$R$ tensor decompositions to $O(2^{n}dR)$ size.
    We evaluate both methods in Section \ref{sec:eval}.

    \subsection{Neural Networks}
    Neural networks express execution time as a nonlinear composition of weighted summations, and thereby circumvent predictor selection and modeling domain discretization.
    Feed-forward multi-layer perceptrons (MLP) were among the first neural network architectures evaluated for performance modeling~\cite{ipek2006efficiently,lee2007methods}.
    The architectures of these models consist of stacks of layers, each containing an array of units characterized by weights and an activation function.
    Subsequent studies have compared the prediction accuracy of MLPs against other black-box methods~\cite{malakar2018benchmarking} and explored their use as a cost model to guide optimal parameter selection~\cite{tillet2017input}.
    Optimization of MLPs using training data at distinct process counts has been shown to facilitate incremental extrapolation for large-scale scientific simulations~\cite{marathe2017performance}.
    MLPs have additionally been used to generate embeddings (i.e., fixed-size vectors) for variable-length features (e.g., instructions within loop nests) extracted from source code as inputs to more sophisticated architectures.
    Some of these architectures include recurrent and/or recursive neural networks, which have been evaluated to predict throughput of a block of instructions~\cite{mendis2019ithemal,sykora2022granite}, deep-learning pipelines~\cite{steiner2021value}, and to facilitate automatic code generation~\cite{baghdadi2021deep}.
    Additionally, graph neural networks can be formulated from directed acyclic graph representations of compiled source code to predict throughput of tensor-based workloads~\cite{kaufman2021learned}.
    As these models rely on significantly large and un-parameterized training sets, we do not evaluate them.
    We instead evaluate MLP architectures, which themselves feature a large design space, on parameterized training sets.
        
    \subsection{Kernel-based}
    Kernel methods project the modeling domain onto a higher-\newline dimensional space characterized by user-specified kernel functions (e.g., sigmoid).
    Support vector machines (SVM) are global models formulated by partitioning the modeling domain so as to minimize a weighted distance between model predictions and observed performance.
    Gaussian process (GP) regression assumes that observations are samples of a multivariate normal distribution, and are characterized by a mean vector and a positive-definite covariance matrix.
    Although both methods have been evaluated for performance modeling~\cite{malakar2018benchmarking}, they are best-suited when limited training data is available and/or when it is acquired incrementally (e.g., to guide optimal parameter selection~\cite{vuduc2004statistical,duplyakin2016active,menon2020auto}).
    Recent works advance and generalize the use of GPs for learning application performance across independent problems (i.e., multi-task learning) by proposing use of linear combinations of GPs~\cite{liu2021gptune}, piecewise-additive GPs~\cite{luo2021non}, and deep GPs~\cite{luszczek2023combining}.
    We evaluate both SVMs and basic GP regression models in this work.

    \subsection{Recursive Partitioning}
    Decision trees partition the modeling domain into hyper-rectangles, each of which is assigned a constant value that models the execution time of configurations mapped within its sub-domain.
    These hyper-rectangles are constructed by recursively splitting a parameter's range to minimize a loss function (e.g., MSE) on a subset of observations.
    Random forest regression models execution time as an average of these constants;
    each tree is constructed using a random sample of observations via bootstrap aggregation, while each tree's splits are constructed by minimizing a loss on a random sample of a subset of observations.
    Extremely-randomized tree regression, which has been shown to be among the most accurate methods for performance modeling~\cite{jain2013predicting,marathe2017performance,malakar2018benchmarking,ibeid2019learning},
    computes splits randomly instead of minimizing a specified error metric.
    Gradient-boosting regression constructs trees sequentially by fitting residuals attained by the previously-constructed sequence of trees, which are proportional to the negative gradients of the chosen loss function.
    These methods have each been evaluated to construct surrogate models within a Bayesian optimization framework to guide optimal parameter selection~\cite{wu2022autotuning,wu2023ytopt}.
    In our experimental evaluation, we tune over a subset of hyper-parameters shared by each method, including forest size (i.e., number of trees) and tree-depth.
    
    \subsection{Instance-based}
    Instance-based methods construct models on-the-fly using execution times of observed configurations.
    We evaluate the $k$-nearest neighbors method, which predicts execution time by interpolating the execution times of the $k$ nearest observed configurations relative to a specified distance metric, and which has been previously evaluated for modeling application performance~\cite{malakar2018benchmarking}.

\section{Tensor Completion}\label{sec:tc}
    Tensor completion is the task of building a model to approximate a tensor based on a subset of observed entries~\cite{acar2011scalable,liu2012tensor}. 
    A low-rank tensor factorization is typically used to model a partially-observed tensor, which is optimized via tensor completion.
    We review models and optimization methods for tensor completion in this section.
    \subsection{Low-Rank Tensor Factorization}
        A tensor $\mathcal{T}\in\mathcal{R}^{I_{1}\times\cdots\times I_{d}}$ has order $d$ (i.e., $d$ modes/indices), dimensions $I_{1}$-by-$\ldots$-by-$I_{d}$, and elements $t_{i_{1},\ldots,i_{d}}\equiv t_{\mathbf{i}}$ where $\mathbf{i}\in\otimes_{j=1}^{d}\{1,\ldots,I_{j}\}$.
        A tensor exhibiting low-rank structure can be approximated as a summation of tensor products (i.e., canonical-polyadic decomposition~\cite{hitchcock1927expression}) or using other tensor factorizations such as Tucker~\cite{kolda2009tensor}.
        The canonical-polyadic (CP) decomposition of an order three tensor $\mathcal{T}\in\mathcal{R}^{I_{1}\times I_{2}\times I_{3}}$ has the form
        \begin{align}\label{eq:tccp}
        t_{i_{1},i_{2},i_{3}}\approx\sum_{r=1}^{R}u_{i_{1},r}v_{i_{2},r}w_{i_{3},r}\equiv \hat{t}_{i_{1},i_{2},i_{3}},
        \end{align}
        where $R$ denotes the rank of the decomposition, and $\mathbf{U}\in\mathcal{R}^{I_{1}\times R}$, $\mathbf{V}\in\mathcal{R}^{I_{2}\times R}$, $\mathbf{W}\in\mathcal{R}^{I_{3}\times R}$ are factor matrices.
        The size of a CP decomposition grows linearly with tensor order for fixed dimensions and fixed CP rank, and linearly with CP rank for a fixed order and fixed dimensions.
        Tensor completion is an effective technique to attain a low-rank CP decomposition when only a subset of tensor elements are observed~\cite{tomasi2005parafac,acar2011scalable}.
    
        A CP decomposition optimized via tensor completion should accurately represent observed tensor entries, yet generalize effectively to unobserved entries.
        The set of observed entries of an order-$3$ tensor $\mathcal{T}$ is represented by an index set $\Omega\subseteq \{1,\ldots,I_{1}\}\times\{1,\ldots,I_{2}\}\times\{1,\ldots,I_{3}\}$ so that every element $\mathbf{i}\equiv (i_{1},i_{2},i_{3})\in\Omega$ has an associated observation $t_{\mathbf{i}}\equiv t_{i_{1},i_{2},i_{3}}$.
        Tensor completion involves minimizing an objective function that consists of a convex loss function and regularization terms in each factor matrix.
        Regularization decreases model parameter variance and thereby avoids over-fitting to observed entries.
        The choice of element-wise loss function $\phi$ reflects both the distribution of observed elements and the selected error metric.
        Given an observed dataset $\Omega$, CP rank $R$, and regularization parameter $\lambda$, methods for tensor completion with CP decomposition of an order-3 tensor minimize the objective function $g(U,V,W)=$
        \begin{align}\label{eq:general_obj}
            \lambda(||\mathbf{U}||_{F}^{2}+||\mathbf{V}||_{F}^{2}+||\mathbf{W}||_{F}^{2})+\sum_{(i_{1},i_{2},i_{3})\in \Omega} \phi(t_{i_{1},i_{2},i_{3}},\hat{t}_{i_{1},i_{2},i_{3}}).
        \end{align}
        Minimization of Equation \ref{eq:general_obj} using least-squares loss functions $\newline\phi(t_{i_{1},i_{2},i_{3}},\hat{t}_{i_{1},i_{2},i_{3}})=(t_{i_{1},i_{2},i_{3}}-\hat{t}_{i_{1},i_{2},i_{3}})^{2}$ and general convex loss functions $\phi$ (i.e., generalized tensor completion) were initially studied by ~\cite{tomasi2005parafac,acar2011scalable} and ~\cite{hong2020generalized}, respectively.
        We apply tensor completion to fit a CP decomposition to a partially-observed tensor representing benchmark execution times by minimizing both least-squares and general convex loss functions. We leave exploration of alternative tensor decompositions to future work.

    \subsection{Optimization Methods}
        Both exact and approximate CP decomposition of fully- or partially-observed tensors is an NP hard problem~\cite{hillar2013most}.
        We summarize numerical methods for tensor completion that minimize Equation \ref{eq:general_obj} instantiated with various loss functions $\phi$ and any tensor order $d\ge 3$.

        \subsubsection{Least-squares loss}\label{sec:tc:llsf}
        Methods for tensor completion typically fit a CP decomposition to a partially-observed tensor by minimizing Equation \ref{eq:general_obj} with the least-squares loss function  $\phi(t_{i_{1},i_{2},i_{3}},\hat{t}_{i_{1},i_{2},i_{3}})=(t_{i_{1},i_{2},i_{3}}-\hat{t}_{i_{1},i_{2},i_{3}})^{2}$.
        The Levenberg-Marquardt method and the alternating least-squares method (ALS) were the first methods proposed and evaluated for this problem by ~\cite{tomasi2005parafac} and ~\cite{acar2011scalable}, respectively.
        ALS, which we use in Section \ref{sec:cpd_model_formulation:interp}, sweeps over the rows of each factor matrix, fixing the elements of all but one and minimizing, over the $i_{1}$'th row of $\mathbf{U}$, the objective
        \begin{align*}
            g(\mathbf{u_{i_{1}}})=\frac{1}{|\Omega_{i_{1}}|}\bigg[\sum_{(i_{2},i_{3})\in \Omega_{i_{1}}} (t_{i_{1},i_{2},i_{3}}-\hat{t}_{i_{1},i_{2},i_{3}})^{2}\bigg]+\lambda||\mathbf{u}_{i_{1}}||_{2}^{2},
        \end{align*}
        where $(i_{2},i_{3})\in\Omega_{i_{1}}$ if and only if $(i_{1},i_{2},i_{3})\in\Omega$.
        Minimization of each row-wise convex objective involves 
        solving a linear least-squares problem.
        The total arithmetic complexity of ALS is \linebreak $\mathcal{O}((\sum_{j=1}^{d}I_{j})R^{3}+|\Omega|dR^{2})$~\cite{singh2022distributed}.
        This overhead may be circumvented by optimizing factor matrix elements individually.
        In particular, the cyclic coordinate descent method (CCD), which was first proposed by ~\cite{shin2014distributed} and later by ~\cite{karlsson2016parallel} to enhance scalability in distributed-memory settings, 
        minimizes the scalar objective $g(u_{i,r})$ and thereby reduces the arithmetic complexity of ALS by a factor of $R$.
        Both ALS and CCD achieve monotonic convergence,
        yet CCD often exhibits slower convergence due to its decoupling of row-wise factor matrix updates.
        Stochastic gradient descent instead iteratively updates all factor matrix elements at once and calculates the partial derivative of the corresponding objective using a random subset of observations within $\Omega$.
        The efficiency of each method has been explored for tensor completion in both shared and distributed-memory settings~\cite{smith2016exploration}.
        
        \subsubsection{Generalized loss functions}\label{sec:tc:glf}
        Tensor completion may fit CP decompositions to
        partially-observed tensors by minimizing alternative loss functions~\cite{hong2020generalized}.
        These loss functions often indirectly apply constraints to elements within factor matrices (e.g., non-negativity), which are enforced by incorporating additional terms (e.g., penalty and barrier functions) into the objective.
        Alternating minimization and coordinate minimization may apply Newton's method to optimize row-wise and scalar nonlinear subproblems, respectively.
        In particular, alternating minimization via Newton's method (AMN) updates initial row-vector iterates $\mathbf{u}_{i_{1}}^{(0)}$ as follows until convergence:
        \begin{align}\label{eq:glf:row}
            \mathbf{u}_{i_{1}}^{(l+1)}&\gets \mathbf{u}_{i_{1}}^{(l)} - \mathbf{H}_{f}^{-1}(\mathbf{u}_{i_{1}}^{(l)})\nabla g(\mathbf{u}_{i_{1}}^{(l)}),
        \end{align}
        where $\nabla g(\mathbf{u}_{i_{1}}^{(l)}) = \sum_{(i_{2},i_{3})\in\Omega_{i}} \nabla \phi(\mathbf{u}_{i_{1}}^{(l)}) + 2\lambda \mathbf{u}_{i_{1}}^{(l)},\mathbf{H}_{f}(\mathbf{u}_{i_{1}}^{(l)})=\linebreak\sum_{(i_{2},i_{3})\in \Omega_{i}} \mathbf{H}_{\phi}(\mathbf{u}_{i_{1}}^{(l)}) + 2\lambda \mathbf{I}$, and $\mathbf{H}_{f},\mathbf{H}_{\phi}$ are Hessian matrices.
        Gradient-based methods such as stochastic gradient descent and L-BFGS have also been proposed~\cite{hong2020generalized}, as have 
        Gauss-Newton and quasi-Newton methods that optimize all factor matrices with each iteration~\cite{singh2022distributed}.

        We apply generalized tensor completion in Section \ref{sec:cpd_model_formulation:extrap} to achieve positive factor matrices by minimizing the MLogQ2 error metric (see Section \ref{sec:problem_statement:assessment}).
        We append Equation \ref{eq:general_obj} with log barrier functions applied element-wise to each factor matrix;
        each term is scaled by barrier parameter $\eta$.
        We then adapt the AMN method to solve a sequence of row-wise subproblems following Equation \ref{eq:glf:row}, each for a fixed value $\eta$, and decrease $\eta$ geometrically until it is smaller than regularization parameter $\lambda$.
        This procedure follows established interior-point methods for nonlinear optimization~\cite{wright1999numerical}.

\section{Application Performance Modeling via Tensor Completion}\label{sec:low-rank-perf-model}

    We apply tensor completion to optimize a low-rank canonical-polyadic (CP) decomposition of a tensor representation of an application's execution time.
    Our modeling framework enables user-specified partitioning of the modeling domain and provides loss functions to minimize user-specified error metrics.
    We formulate our approach to address the modeling problem described in Section \ref{sec:problem_statement:overall}.

    \begin{figure}[t] 
      \centering
      \includegraphics[width=1\linewidth]{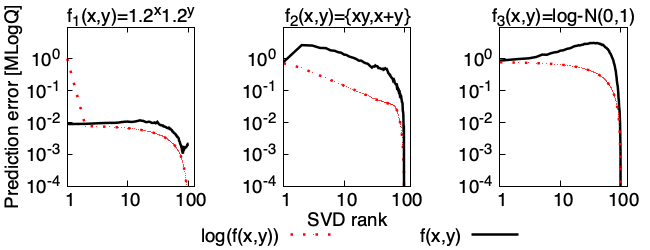} 
      \caption{Singular value decompositions of three discretized functions evaluated for $1\le x,y\le 100$. The two behaviors of $f_{2}$ are split along $x+y\le 100$. Each element of $f_{1},f_{2}$ is multiplied by $(1+\mathcal{N}(0,.01))$, where $\mathcal{N}(\mu,\sigma)$ denotes a sample from a  normal distribution.
      }
      \label{plot:synthetic}
  \end{figure}
  \begin{figure}[t]
    \centering
    \includegraphics[width=1\linewidth]{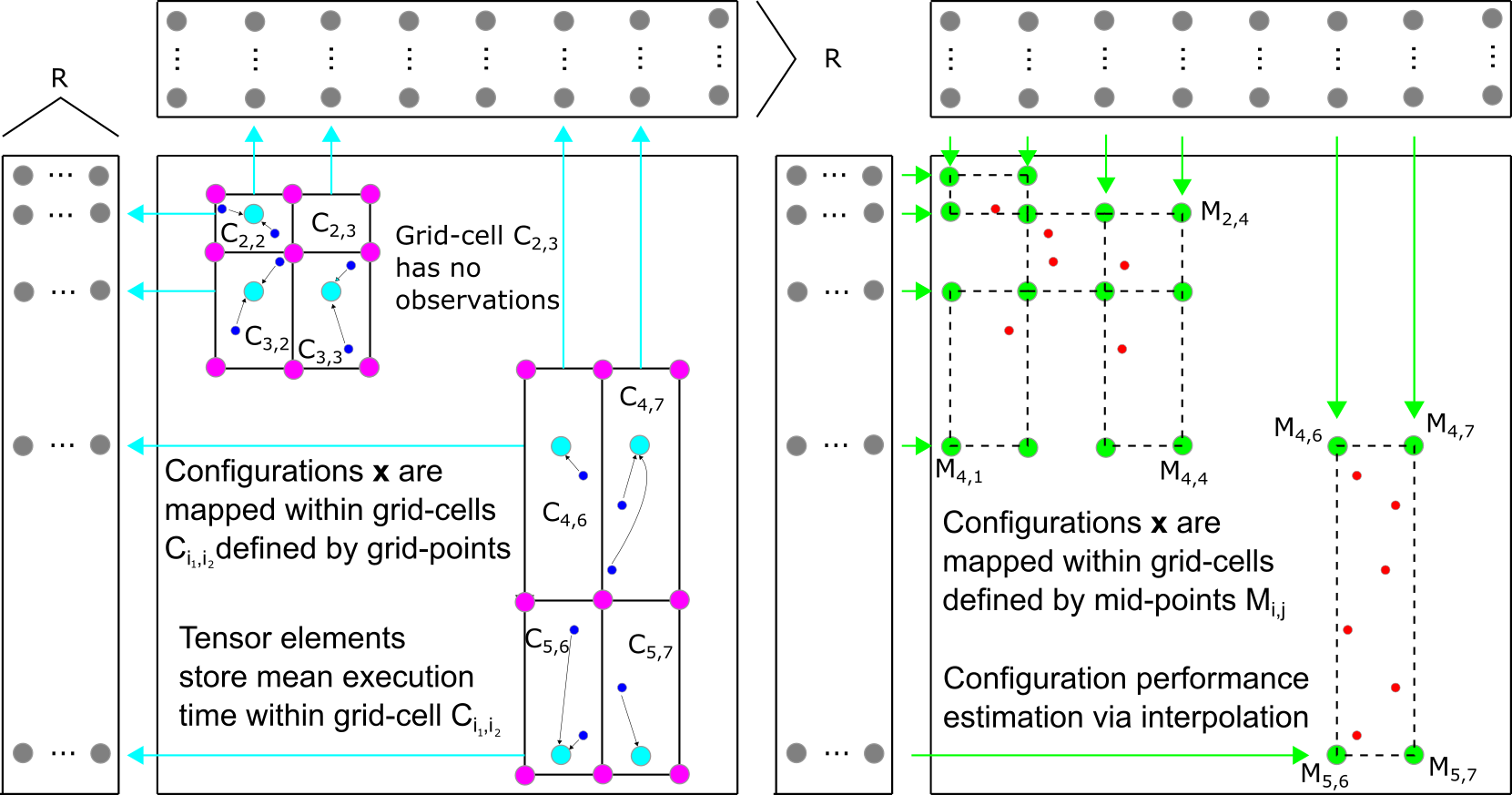}
    \caption[cpd diagram]{Training/inference (left/right) using a rank-$R$ CP decomposition \tikz\draw[gray,fill=gray] (0,0) circle (.5ex);. Tensor elements \tikz\draw[cyan,fill=cyan] (0,0) circle (.5ex); store the sample mean performance of intra-cell training configurations \tikz\draw[blue,fill=blue] (0,0) circle (.5ex);.
    Interpolation of approximate tensor elements \tikz\draw[green,fill=green] (0,0) circle (.5ex); models performance of test configurations \tikz\draw[red,fill=red] (0,0) circle (.5ex);.}
    \label{diagram:grid_space:cpd}
  \end{figure}

    \subsection{Tensor Model Formulation}\label{sec:tensor_model_formulation}
    We use a tensor $\mathcal{T}$ with $d$ dimensions to model execution times of an application with $d$ benchmark parameters.
    Each parameter corresponds to a distinct dimension (mode) of the tensor.
    For parameters $j$ with numerical values (e.g., matrix dimension), we discretize a subset of its range $\tilde{X}^{(j)}\subseteq X^{(j)}$ into $I_{j}$ sub-intervals defined by parameter values $\tilde{X}^{(j)}_{0},\tilde{X}^{(j)}_{1},\ldots,\tilde{X}^{(j)}_{I_{j}}$ using uniform or logarithmic spacing.
    In the case of categorical parameters, we simply index the choices along the corresponding tensor dimension.
    We then define a grid/tensor whose grid-points/entries are associated with the tensor product of both mid-points and end-points of each numerical parameter sub-interval, and distinct choices of each categorical parameter.
    As interpolation is not relevant along tensor modes corresponding to categorical parameters, we hereafter formulate our tensor model by solely considering numerical parameters.
    
    For $d$-parameter configurations, tensor element $t_{i_1,\ldots i_d}\in\mathcal{T}$ is associated with the mid-points of a tensor product of sub-intervals
    $[\tilde{X}^{(1)}_{i_1},\tilde{X}^{(1)}_{i_1+1}],\ldots, [\tilde{X}^{(d)}_{i_d},\tilde{X}^{(d)}_{i_d+1}]$.
    These bounding values define a grid-cell $C_{i_{1},\ldots,i_{d}}$ with mid-point $(M^{(1)}_{i_1},\ldots,M^{(d)}_{i_d})$, which we associate with element $t_{i_1,\ldots i_d}$.
    If the $j$'th parameter is discretized with logarithmic spacing, the mid-point of any sub-interval is given by $M^{(j)}_{i_j}=\lceil e^{(\log(X^{(j)}_{i_j}) + \log(X^{(j)}_{i_j+1}))/2}\rceil$.
    Given a set of executed configurations, $t_{i_1,\ldots, i_d}$ stores the mean execution time among those mapped within cell $C_{i_{1},\ldots,i_{d}}$.
    We leave evaluation of alternative quadrature schemes for future work.
    
    Estimates are available for all tensor elements following completion of the tensor.
    We therefore use linear interpolation to predict execution times for arbitrary configurations $\mathbf{x}\in\tilde{\mathcal{X}}$.
    Given such a configuration, let $\hat{t}_{\mathbf{i}}\equiv \hat{t}_{i_1,\ldots, i_d}$ be an estimate of the tensor element $t_{\mathbf{i}}$ associated with the grid-point $M_{\mathbf{i}}\equiv(M^{(1)}_{i_1},\ldots,M^{(d)}_{i_d})$ just below $\mathbf x$, so each $x_j\in [M^{(j)}_{i_j},M^{(j)}_{i_j+1})$.
    Then, using $h_j(x)=x$ for uniformly and $h_j(x)=\log (x)$ for logarithmically discretized parameter $j$, linear interpolation of neighboring tensor elements gives the execution time prediction,
    \begin{align}\label{eq:dense_tensor_model}
        \sum_{\mathbf{a}\in\{0,1\}^{d}}\hat{t}_{\mathbf{i}+\mathbf{a}}\cdot\prod_{j=1}^{d}\left(1- \frac{|h_j(x_{j})-h_j(M^{(j)}_{\mathbf{i}+\mathbf{a}})|}{h_j(M^{(j)}_{\mathbf{i}+\mathbf{e}_{j}})-h_j(M^{(j)}_{\mathbf{i}})}\right).
    \end{align}
    If $x_j\in [X^{(j)}_{0},M^{(j)}_{0})$ or $x_j\in [M^{(j)}_{I_{j}-1},X^{(j)}_{I_{j}}]$, we instead apply linear extrapolation along the $j$'th mode using the corresponding values. 
    If $x_j\notin[X^{(j)}_{0},X^{(j)}_{I_{j}}]$ and thus $\mathbf{x}\notin\tilde{\mathcal{X}}$, we forgo both interpolation and linear extrapolation along the $j$'th mode and instead propose an alternative technique for the estimation of tensor elements, which we describe in Section \ref{sec:cpd_model_formulation:extrap}.

    The selection of grid-spacing and sub-interval size along each dimension should facilitate accurate compression of tensor $\mathcal{T}$ with small CP rank.
    Our modeling framework therefore enables user-directed domain discretization via specification of uniform or logarithmic to balance interpolation error and low-rank approximation error.
    We next describe two techniques to attain estimates for all elements of tensor $\mathcal{T}$ via low-rank CP decomposition.
    Each targets scale-independent minimization of prediction error across the modeling domain as described in Section \ref{sec:problem_statement:assessment}.

    \subsection{CP Decomposition Model Formulation for Interpolation}\label{sec:cpd_model_formulation:interp}
    We describe a technique to optimize a low-rank CP decomposition of our tensor model that is both accurate and efficient in the interpolation setting described in Section \ref{sec:problem_statement:overall}.
    First, we transform observed entries to enable the use of a loss function that applies the MSE aggregate error metric, which we find is most efficient and least susceptible to round-off error among all error metrics summarized in Section \ref{sec:problem_statement:assessment}.
    Specifically, we minimize Equation \ref{eq:general_obj} with $\phi(t_{\mathbf{i}},\hat{t}_{\mathbf{i}})=(\log(t_{\mathbf{i}})-\hat{t}_{\mathbf{i}})^{2}$ using the ALS method described in Section \ref{sec:tc:llsf}.
    This simple technique minimizes deviation in the scale of observed execution times and thereby enhances the robustness of MSE to both noise and biased under/over-prediction.
    This is observed in Figure \ref{plot:synthetic} using singular value decompositions (SVD), which minimize $\sqrt{\text{MSE}}$.
    In particular, the log-transformed matrices achieve a monotonic decrease in MLogQ prediction error with increasing SVD rank,
    whereas the same error metric can increase with increasing SVD rank for untransformed matrices.
    We assign non-positive entries a value 
    $10^{-16}$ prior to evaluation of MLogQ for all matrices in Figure \ref{plot:synthetic}.

    Use of logarithmic transformations attains non-negative CP decomposition model output implicitly without the requisite constraints to enforce non-negativity in factor matrix elements.
    The resulting model approximates the execution time $f(\mathbf{x})$ of configuration $\mathbf{x}$ as
    \begin{align*}
        m(\mathbf{x})=\sum_{\mathbf{a}\in\{0,1\}^{d}}{\rm e}^{\hat{t}_{\mathbf{i}+\mathbf{a}}}\cdot\prod_{j=1}^{d}\left(1- \frac{|\log(x_{j})-\log(M^{j}_{\mathbf{i}+\mathbf{a}})|}{\log(M^{j}_{\mathbf{i}+\mathbf{e}_{j}})-\log(M^{j}_{\mathbf{i}})}\right),
    \end{align*}
    where
    $M_{\mathbf{i}}^{j}\le x_{j}< M_{\mathbf{i}+\mathbf{e}_{j}}^{j},\forall j\in\{1,\ldots,d\}$.
    As mentioned in Section \ref{sec:tensor_model_formulation}, if $x_j\in [X^{(j)}_{0},M^{(j)}_{0})$ or $x_j\in [M^{(j)}_{I_{j}-1},X^{(j)}_{I_{j}}]$, we instead apply linear extrapolation along the $j$'th mode using the corresponding values.
    We summarize training and inference in an interpolation setting across a two-dimensional domain in Figure \ref{diagram:grid_space:cpd}.

    \begin{table*}[t]
        \centering
        \begin{tabular}{||m{5em}|m{15em}|m{8em}|m{7em}|m{15em}||} 
            \hline
            Library & \textbf{Application} [parameter description]& Input & Architectural & Configuration \\ 
            \hline\hline
            ExaFMM\cite{yokota2013fmm} & \textbf{Fast multipole method} [number of particles per node $n$, expansion order $ord$, particles per leaf $ppl$, partitioning tree level $tl$] & $2^{12}\le n\le 2^{16}$, $4\le ord\le 15$ & $1\le tpp\le 64$, $1\le ppn\le 64$ & $32\le ppl\le 256$, $0\le tl\le 4$\\
            \hline
            AMG\cite{richards2018quantitative} & \textbf{Algebraic multigrid} [Problem size per process \textit{nx/ny/nz}, coarsening/relaxation/interpolator type \textit{ct}/\textit{rt}/\textit{it}]  & $2^{3}\le \textit{nx,ny,nz}\le 2^{7}$ & $1\le tpp\le 64$, $1\le ppn\le 64$ & $\textit{ct}\in\{0,3,6,8,10,21,22\}$, $\textit{rt}\in\{0,3,4,6,8,13,14,16,17,18\}$, $\textit{it}\in\{0,2,3,4,5,6,8,9,12,13,14,16,17,18\}$\\
            \hline
            KRIPKE\cite{kunen2015kripke} & \textbf{Discrete ordinates transport} [\textit{groups}, \textit{legendre}, \textit{quad}, \textit{dset}, \textit{gset}, layout \textit{l}, solver \textit{s}]  & $2^{3}\le \textit{groups}\le 2^{7}$, $0\le \textit{legendre}\le 5$, $2^{3}\le \textit{quad}\le 2^{7}$ & $1\le tpp\le 64$, $1\le ppn\le 64$ & $\textit{l}\in\{dgz,dzg,gdz,gzd,zdg,zgd\}$, $s\in\{sweep,bj\}$, $8\le \textit{dset}\le 64$, $1\le \textit{gset}\le 32$\\
            \hline
        \end{tabular}
        \caption{Parameter space description for algebraic multigrid and multiple parallel scientific applications. Parameters \textit{N}, \textit{ppn} and \textit{tpp} refer to node count, processes-per-node and threads-per-process under the constraint that $64\le ppn\cdot tpp\le 128$.}
        \label{parameter_table}
    \end{table*}

    \subsection{CP Decomposition Model Formulation for Extrapolation}\label{sec:cpd_model_formulation:extrap}
    We next describe a technique to optimize a low-rank CP decomposition of our tensor model that enables inference in the extrapolation setting described in Section \ref{sec:problem_statement:overall}.
    A CP decomposition (e.g., that formulated in Section \ref{sec:cpd_model_formulation:interp}), appears difficult to use in this setting, as one must predict unseen rows of its factor matrices, which may lack structure due to cancellation of its positive and negative elements.
    However, a rank-1 CP decomposition exhibits structure naturally, as performance typically increases or decreases monotonically as the value of a numerical parameter increases.
    One can then simply extrapolate a sequence using line-fitting techniques.
    To retain accuracy in CP factor matrices that correspond to parameters that are not being extrapolated, we do not restrict CP rank to 1 and instead apply rank-1 factorizations to strictly positive factor matrices.
    We then apply line-fitting to extrapolate each rank-1 factorization.

    We utilize the generalized tensor completion framework summarized in Section \ref{sec:tc:glf} to ensure positive factor matrices.
    Specifically, we minimize Equation \ref{eq:general_obj} with loss function $\phi(t_{\mathbf{i}},\hat{t}_{\mathbf{i}})=(\log(t_{\mathbf{i}})-\log(\hat{t}_{\mathbf{i}}))^{2}$ using the interior-point method described in Section \ref{sec:tc:glf}.
    We then leverage the Perron-Frobenius theorem, which states that the  best rank-1 approximation to
    a strictly positive matrix is itself positive, ensuring positive predictions.
    While positive factor matrices attained in this way are not typically low-rank, we find empirically that this technique retains sufficient accuracy because the low-rank approximation error does not compromise the contributions from other factor matrices when estimating a tensor element.
    We leave evaluation of alternative schemes to enable robust extrapolation using a CP decomposition for future work.

    We fit a univariate spline model to the positive left singular vector of each factor matrix corresponding to a numerical or architectural benchmark parameter.
    In particular, we use elements of this vector as a training set, which we log-transform element-wise, to configure a MARS model, which we summarized in Section \ref{sec:review:grid}.
    Thus, for $d=3$ yet without loss of generality, when presented a configuration $\mathbf{x}\notin\tilde{\mathcal{X}}$ such that $x_{1}\notin[X^{(1)}_{0},X^{(1)}_{I_{1}}]$, $x_{2}\in[X^{(2)}_{0},X^{(2)}_{I_{2}}]$, $x_{3}\in[X^{(3)}_{0},X^{(3)}_{I_{3}}]$ we modify 
    Equation \ref{eq:tccp} by estimating tensor element $t_{\mathbf{i}}$ as
    \begin{align*}
      \hat{t}_{\mathbf{i}}=\sum_{r=1}^{R}e^{\hat{m}(\log(x_{1}))}\hat{\sigma}\hat{v}_{r}v_{i_{2},r}w_{i_{3},r},
    \end{align*}
      where $\mathbf{U}\approx\hat{\mathbf{u}}\hat{\sigma}\hat{\mathbf{v}}^{T}$ and $\hat{m}:\mathbb{R}\rightarrow\mathbb{R}$ is the spline model configured by MARS using training set $\log(\hat{\mathbf{u}})$.
    In a general setting with $d\ge 2$, when presented with a configuration $\mathbf{x}\notin\tilde{\mathcal{X}}$ such that interpolation is invoked along at least one numerical benchmark parameter (i.e., $\exists j$ s.t. $x_{j}\in[X^{(j)}_{0},X^{(j)}_{I_{j}}]$), we utilize Equation \ref{eq:dense_tensor_model} to approximate execution time $f(\mathbf{x})$ and treat extrapolated numerical parameters as we do categorical (i.e., no interpolation along its tensor dimension).

\section{Experimental Methodology}\label{sec:exp_methodology}

    \subsubsection{Machine}
        We use the Stampede2 supercomputer at Texas Advanced Computing Center (TACC)\cite{Stanzione:2017:SEX:3093338.3093385}.
        Stampede2 consists of 4200 Intel Knights Landing (KNL) compute nodes, each connected by an Intel Omni-Path (OPA) network with a fat-tree topology.
        Each KNL compute node provides 68 cores with 4 hardware threads per core.
        We use Intel environment 18.0.2, Intel MPI environment 18.0.2, and corresponding icc, mpicc, and mpicxx compilers.
        We generate all models sequentially.

    \begin{figure*}[t] 
        \centering
        \includegraphics[width=1\linewidth]{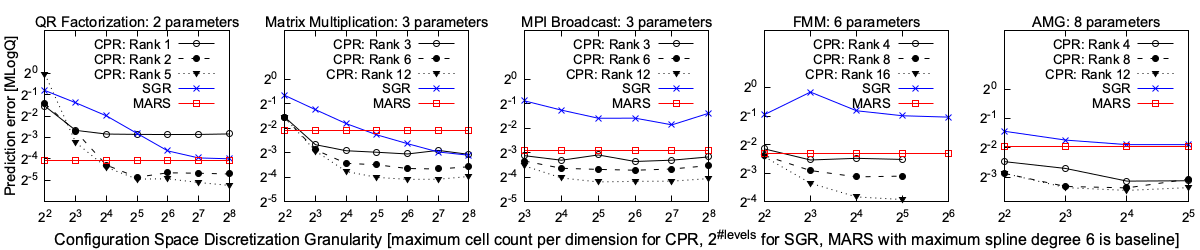} 
        \caption{Prediction accuracy for grid-based models. Discretization granularity signifies the number of cells for CPR and $2^{\text{discretization level}}$ for SGR.
        The training-set sizes for each kernel/application are $2^{16},2^{16},2^{15},2^{15},2^{14}$, respectively.
        }
        \label{plot:error_vs_disc} 
    \end{figure*}
    \begin{figure*}[t] 
        \centering
        \includegraphics[width=1\linewidth]{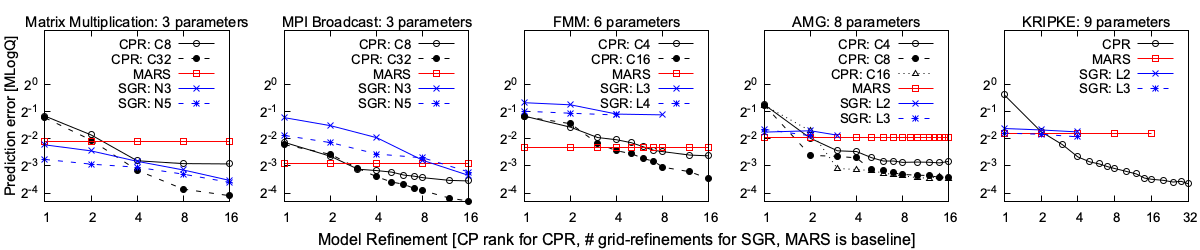} 
        \caption{Prediction accuracy for grid-based models following refinement of initial model (CP rank for CPR, sparse grid for SGR). C$k$ and L$k$ denote cell count and sparse grid-level $k$, respectively. The training-set sizes for each kernel/application are $2^{16},2^{15},2^{15},2^{14},2^{14}$, respectively.}
        \label{plot:error_vs_refinement} 
    \end{figure*}
    \begin{figure*}[t] 
        \centering
        \includegraphics[width=1\linewidth]{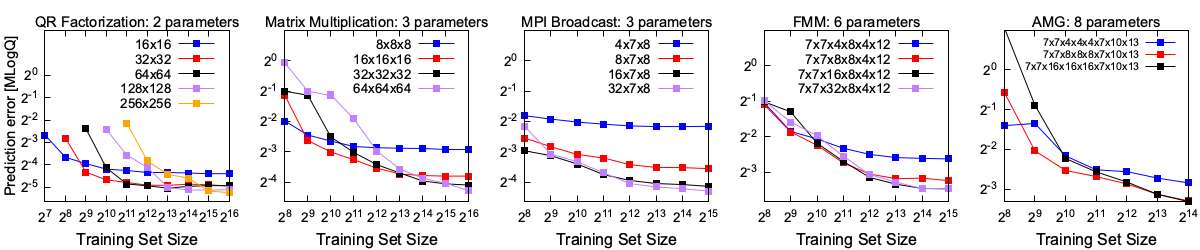}
        \caption{Prediction accuracy for the CPR model. Tensors corresponding to each CPR model becomes increasingly dense with increasing training set size. For each tensor, the minimal prediction error across CP ranks is reported.}
        \label{plot:error_vs_completion_density} 
    \end{figure*}

    \subsubsection{Application Libraries}
        We execute Intel MKL's GEMM (MM) routine $C_{m\times n}\gets A_{m\times k}B_{k\times n} + \beta C_{m\times n}$ in a single-threaded environment with matrix dimensions $32\le\textit{m},\textit{n},\textit{k}\le4096$.
        We execute Intel MKL's GEQRF (QR) routine $A_{m\times n}\gets \alpha Q_{m\times n}R_{n\times n}$ in a single-threaded environment with matrix dimensions $32\le\textit{m},\textit{n}\le 262144$ such that $m\ge n$ and all three matrices fit in memory.
        We execute Intel MPI's Broadcast (BC) routine on $\{1,2,4,8,16,32,64,128\}$ nodes with $\{1,2,4,8,16,32,64\}$ processes-per-node (\textit{ppn}), and message size $2^{16}\le\textit{m}\le 2^{26}$.
        ExaFMM\cite{yokota2013fmm} is an open source library for fast multipole methods aimed towards exascale systems.
        AMG\cite{richards2018quantitative} is a parallel algebraic multigrid solver for linear systems arising from problems on unstructured grids.
        Kripke~\cite{kunen2015kripke} is a proxy application for modern discrete ordinates neutral particle transport applications.
        We execute each application configuration on a single-node, and model the execution times of the \textit{m2l\_\&\_p2p} kernel for ExaFMM and the total solve time for AMG and Kripke.
        Table \ref{parameter_table} details each application's parameter space.
    
    \subsubsection{Dataset Generation}
        To optimize and evaluate performance models, we generate training sets and test sets for each kernel and application.
        Individual configurations within both sets are formulated using distinct sampling strategies for distinct parameter types.
        In particular, the specified ranges of both input parameters (e.g., matrix dimension, message size) and architectural parameters (e.g., node count, ppn) are sampled log-uniformly at random, while the ranges of each configuration parameter (e.g., block size) are sampled uniformly at random.
        The test sets for matrix multiplication, QR factorization, MPI broadcast, ExaFMM, AMG, and Kripke feature 1000, 1000, 10484, 2512, 21534, and 8745 configurations, respectively.
        Each kernel configuration is executed 50x or until the coefficient of variation of its corresponding execution time is $<0.01$. 
        Each application configuration is executed once.

    \subsubsection{Model Tuning and Evaluation}\label{sec:exp_methodology:model_tuning}
    
\begin{figure*}[t] 
        \centering
        \includegraphics[width=1\linewidth]{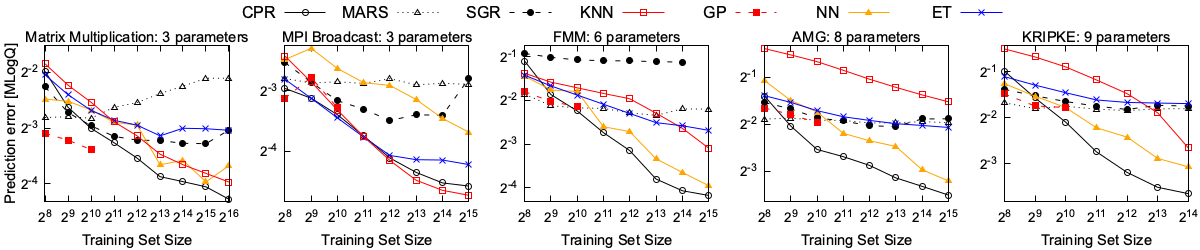} 
        \caption{Prediction error for grid-based models and alternative supervised learning models as a function of training set size. Each datapoint represents the minimum error achieved by exhaustively exploring the corresponding model's hyper-parameters as described in Section \ref{sec:exp_methodology:model_tuning}.}
        \label{plot:error-vs-training_set_size}
    \end{figure*}

\begin{figure*}[t] 
        \centering
        \includegraphics[width=1\linewidth]{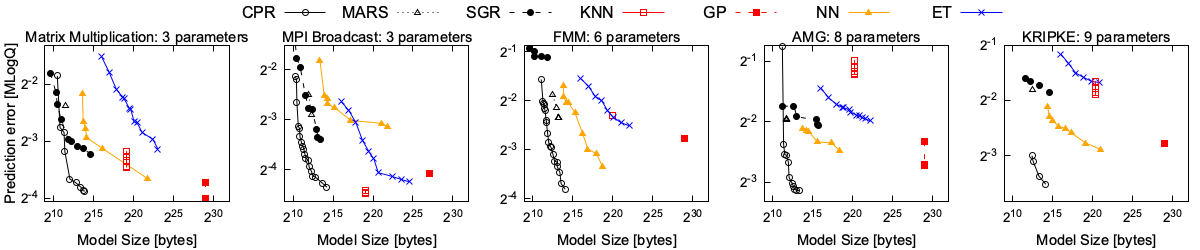} 
        \caption{Prediction error for grid-based models and alternative supervised learning models as a function of model size. The complexity of each model is varied by evaluating all hyper-parameters as described in Section \ref{sec:exp_methodology:model_tuning}. All models are trained using 8192 samples.}
        \label{plot:error-vs-model_size}
    \end{figure*}

        We evaluate the following prediction methods: sparse grid regression (SGR), multi-layer perceptrons (NN), random forests (RF), gradient-boosting (GB), extremely-randomized trees (ET), Gaussian process regression (GP), support vector machines (SVM), adaptive spline regression (MARS), k-nearest neighbors (KNN), and our methodology described in Section \ref{sec:low-rank-perf-model} (CPR).
        We optimize these models using a random sample from each training set and log-transform execution times and application parameters.
        Results for SVM, RF, and GB are not provided in Figures \ref{plot:error-vs-training_set_size}, \ref{plot:error-vs-model_size}, and \ref{plot:error-vs-extrap-env1} because each attains a strictly less accurate model than others of the same model class: GP and ET, respectively.


        We utilize the Cyclops Tensor Framework~\cite{solomonik2014massively} for efficient CP decomposition optimization.
        We utilize the PyEarth library to evaluate MARS, the SG++~\cite{pflueger10spatially} library to evaluate SGR, 
        and the Scikit-Learn~\cite{pedregosa2011scikit} library to evaluate the remaining methods.
        We evaluate all relevant model configurations using the same training set and forgo training via cross-validation.
        In particular, we evaluate CP ranks $1\rightarrow 64$ and grid-cell counts $4\rightarrow 256$ per dimension for CPR; discretization levels $2\rightarrow 8$, $1\rightarrow 16$ refinements, and $4\rightarrow 32$ adaptive grid-points for SGR; max spline degrees of $1\rightarrow 6$ for MARS; max tree depths of $2\rightarrow 16$ and max tree counts of $1\rightarrow 64$ for RFR, GBR, and ETR; $1\rightarrow 6$ neighbors for KNN, \{RationalQuadratic,RBF,DotProduct+WhiteKernel,\linebreak Matern,ConstantKernel\} covariance kernels for GPR; \{poly,rbf\} kernels and polynomial degrees $1\rightarrow 3$ for SVM; $1\rightarrow 8$ hidden layers each of size $2\rightarrow 2048$ and \{relu,tanh\} activation functions for NN.

        CPR places grid-points equally-spaced on a log-scale/linear-scale along the ranges of input,architectural/configuration parameters, and interpolates as necessary.
        We select regularization parameter $\lambda=10^{-k},\forall k\in\{-6,\ldots,-3\}$ for CPR and SGR.
        We set a maximum number of 100 sweeps and 1000 iterations of alternating least-squares and conjugate gradient for CPR and SGR, respectively, and set a tolerance of $10^{-4}$ for SGR.
        We initialize the barrier parameter $\eta$ for extrapolation experiments using CPR to $10$ and decrease geometrically by a factor of 8 until $\eta<=10^{-11}$.
        In these experiments, we set a maximum of $40$ iterations of Newtons method to solve each row-wise optimization problem for a fixed value $\eta$.

        We assess model prediction error on test configurations using the MLogQ error metric as detailed in Section \ref{sec:problem_statement:assessment}.
        We assess model size by writing performance models optimized for the interpolation setting described in Section \ref{sec:problem_statement:overall} to a file via Python's \textit{joblib} package and measuring file size.
        We forgo consideration of models optimized in $\ge 1000$ seconds in Figures \ref{plot:error-vs-training_set_size} and \ref{plot:error-vs-model_size}, or models of size $\ge$ 10 megabytes in Figure \ref{plot:error-vs-training_set_size}.
        As our focus is on evaluating model accuracy, we forgo consideration of model evaluation latency and training time.

    \section{Experimental Evaluation}\label{sec:eval}
    
\begin{figure*}[t]
        \centering
        \includegraphics[width=1\linewidth]{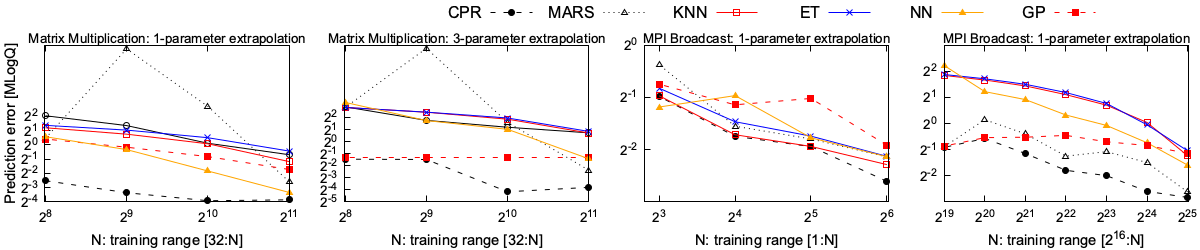} 
        \caption{Extrapolation error for configurations beyond the specified training ranges of matrix dimension $m$ and all matrix dimensions for Matrix Multiplication, and on node count and message size for MPI Broadcast, respectively. All models are trained using 4096 samples.}
        \label{plot:error-vs-extrap-env1}
    \end{figure*}

        We evaluate each class of performance models generated by methods described in Section \ref{sec:review} and specified in Section \ref{sec:exp_methodology:model_tuning}.
        Following the methodology proposed in Section \ref{sec:problem_statement:assessment}, we first assess piecewise/ grid-based models and alternative supervised learning methods for the interpolation problem described in Section \ref{sec:problem_statement:overall}. 
        We then assess each class of models for both single-parameter and multi-parameter extrapolation also described in that section.
        We provide Python scripts\footnote{https://github.com/huttered40/cpr-perf-model/tree/main/reproducibility} to reproduce all experiments summarized in Figures \ref{plot:error_vs_disc}, \ref{plot:error_vs_refinement}, \ref{plot:error_vs_completion_density}, \ref{plot:error-vs-training_set_size}, \ref{plot:error-vs-model_size}, and \ref{plot:error-vs-extrap-env1}.

  \subsection{Model Evaluation for Interpolation}
  
    Existing piecewise/grid-based performance models exhibit fundamental trade-offs between accuracy and scalability.
    Each metric is significantly influenced by discretization geometry (i.e., location of grid-points) and granularity (i.e., number of grid-points dictated by discretization geometry) across the modeling domain. Both influence the observation density within individual grid-cells for a fixed training set.
    We evaluate prediction error in the interpolation setting described in Section \ref{sec:problem_statement:overall} across varying domain discretization and grid refinement strategies.
    We then incorporate alternative supervised learning methods to compare model accuracy as a function of model size and number of training observations.

    \subsubsection{Prediction Accuracy vs. Domain Discretization}
    
    We observe in Figure \ref{plot:error_vs_disc} that compression of high-dimensional regular grids via CP decomposition (CPR) facilitates higher model prediction accuracy than regression across high-dimensional sparse grids (SGR).
    Figure \ref{plot:error_vs_disc} further demonstrates that given sufficiently many training observations, CPR alone among piecewise/grid-based models systematically improves model accuracy by independently varying the discretization granularity along the ranges of numerical parameters across all five benchmarks.
    We remark that when applications feature both numerical and integer/categorical parameters, user-controlled discretization of the modeling domain, which CPR alone facilitates, is critical to formulating an accurate performance model.
    As observed for the FMM and AMG benchmarks in Figure \ref{plot:error_vs_disc}, each of which features both types of parameters, SGR's strategy of varying the discretization granularity across all dimensions simultaneously offers no improvement in prediction accuracy.
    Discretization geometry selected via search alone (MARS) configures global models that are significantly less accurate than CPR.
    For both high-dimensional benchmarks, this enables CPR to achieve up to 4x improvements in mean prediction accuracy over SGR and MARS.

    Figure \ref{plot:error_vs_disc} highlights that use of a regular grid, as opposed to alternatives such as a sparse grid, is most conducive to systematic improvement in model prediction accuracy.
    However, our approach, CPR, is the only available technique to achieve a scalable representation of such a grid for a high-dimensional modeling domain.
    Results in Figures \ref{plot:error_vs_disc} and \ref{plot:error_vs_refinement} show that explicit use of a regular grid isn't necessary;
    even a rank-$4$ CP decomposition model of its corresponding tensor representation is sufficient to achieve higher accuracy than that attained by SGR and MARS.
    Further, the prediction accuracy of CPR continues to improve for a fixed CP rank even as the underlying tensors become increasingly sparse.

      Results in Figure \ref{plot:error_vs_refinement} indicate that CP rank offers the most accurate and efficient mechanism for improving prediction accuracy among those provided by piecewise/grid-based models.
      We observe that as tensor models grow in size, a higher CP rank is needed to leverage the finer discretization of the underlying regular grid.
      However, up to 4x improvements in prediction accuracy are achieved even with CP rank-8 models for the largest evaluated tensors.
      SGR instead refines its underlying sparse grid, each sweep iterating over training data to identify grid-points and corresponding basis functions contributing most to training error.
      The discretization geometry around those grid-points is modified before restarting the procedure on the newly refined grid.
      As observed in Figure \ref{plot:error_vs_refinement}, even with up to 16 grid-refinements (which can take more than 1000 seconds), sparse grid models cannot achieve prediction accuracy competitive with CPR, which does not refine its underlying grid.

    \subsubsection{Prediction Accuracy vs. Training Set Size}
    We observe in Figure \ref{plot:error_vs_completion_density} that CPR improves mean prediction accuracy with increasingly fine regular grids only if the corresponding tensor is sufficiently dense.
    Notably, as the number of benchmark parameters increases, the requisite tensor density to optimize an accurate low-rank CP decomposition model decreases.
    In particular, for MM, a 32x32x32 tensor model offers best prediction accuracy when at least 50\% dense (training set size of $2^{14}$), whereas for AMG, a 7x7x8x8x8x7x10x13 tensor model is $0.07\%$ dense yet most accurate at the same training set size.
    Thus, the size of the training set, along with tensor size, is most indicative of the accuracy these models can attain.

        We tune over the optimal tensor size and CP rank in Figure \ref{plot:error-vs-training_set_size}, which further shows systematic improvements in prediction accuracy with increasingly large training sets.
        CPR achieves the highest prediction accuracy among alternative models using increasingly large training sets of applications FMM, AMG, and KRIPKE in Figure \ref{plot:error-vs-training_set_size}.
        Neural networks are most competitive among alternative models for modeling performance across high-dimensional domains.
        We further observe that accurate CPR models can be optimized with small training sets using small tensors and moderate CP rank.
        However, alternative models are more competitive in this regime, indicating that CPR is most effective relative to the state-of-the-art for training sets of moderate to large size.

    \subsubsection{Prediction Accuracy vs. Model Size}
        We observe in Figure \ref{plot:error-vs-model_size} that CPR models achieve the highest prediction accuracy relative to model size among alternative models for all five benchmarks.
        Improvement in this metric is also observed as the number of application benchmark parameters increases for a fixed-size training set.
        In particular, CPR approximately matches KNN and GP methods in prediction accuracy for the MM and BC kernels, despite using 16384x and 32x less memory, respectively.
        For FMM, AMG, and KRIPKE, the execution times of which depend on at least six parameters, CPR achieves the smallest prediction error outright, and uses 50x less memory than the most accurate neural network model.

        Among piecewise/grid-based models, CPR decouples discretization granularity of the modeling domain from model size if performance across its underlying grids is sufficiently low-rank.
        This is attributed to properties characteristic to CP decomposition and listed in Section \ref{sec:review:grid}.
        Namely, as the number of benchmark parameters increases (a segment of a general trend), the CPD model size grows linearly with tensor order for a fixed rank and tensor size.
        Thus, memory-efficiency is retained even when modeling performance across high-dimensional domains, as we observe in Figure \ref{plot:error-vs-model_size}.
        Fully-connected neural networks, the size of which scales linearly in number of layers, are most competitive with CPR among alternative supervised learning methods for large training sets.
        For all evaluated benchmarks, alternative piecewise/grid-based models (e.g., MARS), while competitive with CPR in model size, cannot fit to complex performance data as accurately using sparse grids or high-degree splines.
        Instance-based methods (e.g., KNN) achieve increasingly poor accuracy in higher dimensions due to increasingly sparse observation of the modeling domain.
        Highly tuned kernel-based methods (e.g., GP) and recursive partitioning methods (e.g., ET) require an exorbitant memory requirement, yet achieve increasingly poor accuracy in higher dimensions.

    \subsection{Model Evaluation for Extrapolation}
        We assess all models for both single-parameter and multi-parameter extrapolation using the matrix multiplication (MM) and MPI broadcast (BC) kernels.
        For each kernel, we select a subset of its numerical or integer parameters and assign the corresponding multi-parameter configurations to a training set or test set according to the magnitudes of the selected parameters.
        Specifically, the test sets for MM consist of all configurations $(m,n,k) : 2048\le m\le 4096$ and all configurations $(m,n,k) : 2048\le m,n,k\le 4096$, while the corresponding training sets consist of all configurations $(m,n,k) : 32\le m< N$ and all configurations $(m,n,k) : 32\le m,n,k< N$ for $2^{8}\le N\le 2^{11}$, respectively.
        The test sets for BC consist of all configurations executed with 128 nodes and all message sizes $m : 2^{25}\le m\le 2^{26}$, while the corresponding training sets consist of all configurations executed with node counts $p : 1\le p\le N$ for $8\le N\le 64$ and all message sizes $m : 2^{16}\le m < N$ for $2^{19}\le N\le 2^{25}$, respectively.
        As we have already ablated these models for the interpolation setting, we randomly select 4096 samples from each training set and report the most accurate model as $N$ is decreased geometrically.

        We observe in Figure \ref{plot:error-vs-extrap-env1} that our approach, CPR, is significantly more accurate than alternative models when extrapolating numerical parameters of MM and BC kernels.
        All alternative models we evaluate overfit to training observations irrespective of model complexity.
        CPR instead captures monotonically increasing performance behavior observed within training observations along the columns of each corresponding factor matrix to enhance extrapolation accuracy.
        CPR also benefits from user-directed discretization granularity along the dimensions corresponding to extrapolated parameters, as a finer discretization (i.e., larger tensor dimension) enables use of a larger training set to improve optimization of spline models as described in Section \ref{sec:cpd_model_formulation:extrap}.
        As $N$ decreases and observed performance behavior among training configurations along factor matrices is less predictive of performance within test set configurations, CPR's extrapolation error increases accordingly.
        However, when extrapolating a single numerical parameter for both MM and BC kernels, CPR is consistently more accurate as $N$ decreases by ~8x and ~2x, respectively.

        Our model is less accurate when extrapolating integer parameters, which we observe when extrapolating node count for the BC kernel in Figure \ref{plot:error-vs-extrap-env1}.
        This can be attributed to the small training set with which the spline model of the corresponding factor matrix's left singular vector is optimized, as well as non-monotonic performance behavior that CPR captures which is characteristic to smaller-scale experiments but not to larger-scale.
        In particular, for this kernel and $N\le 32$, CPR matches KNN in achieving 25\% mean prediction error.
        Overall, these results suggest that if performance behavior of moderate-dimension kernels at small-scale is conducive to that at large-scale, our methodology described in Section \ref{sec:cpd_model_formulation:extrap} is  relatively well-equipped to extrapolate performance accurately.

\section{Conclusion}
  Our experimental analysis demonstrates that a CP decomposition can accurately and efficiently model application performance across high-dimensional parameter spaces.
  These models are significantly more accurate than alternative models relative to model size (and increasingly so as the number of benchmark parameters increases), and systematically reduce prediction error given increasingly many observations of a fixed modeling domain.
  The impetus for tensor completion driven by artificial intelligence (e.g., recommendation systems, image recognition) ensures that performance modeling frameworks that adopt this technique will benefit from its continued optimization and dissemination among software libraries.

  This work presents promising avenues for autotuning research involving the use of tensor factorizations as surrogate models of application performance.
  Remaining challenges using tensor completion in this setting include handling performance observation datasets with different (non-random) structure that reflects exploration and exploitation sampling methods and/or highly-constrained parameter spaces, as well as incorporating methods for efficiently updating CP decompositions to effectively model streaming data in online settings~\cite{mardani2015subspace,kasai2016online,song2017multi}.
  An additional gap in our evaluation for this setting includes optimization of tensor factorizations to target accurate identification of fast configurations (e.g., that proposed in \cite{marathe2017performance}) as opposed to minimizing prediction error over all configurations in aggregate.

\begin{acks}
The first author would like to acknowledge the Department of Energy (DOE) and Krell Institute for support via the DOE Computational Science Graduate Fellowship (grant No.\ DE-SC0019323).
This work used Stampede2 at the Texas Advanced Computing Center through allocation CCR180006 from the Advanced Cyberinfrastructure Coordination Ecosystem: Services \& Support (ACCESS) program, which is supported by National Science Foundation grants \#2138259, \#2138286, \#2138307, \#2137603, and \#2138296.
This research has also been supported by funding from the National Science Foundation via grant \#1942995.
\end{acks}

\bibliographystyle{ACM-Reference-Format}
\bibliography{paper}


\end{document}